\newcommand\apjcls{1}
\newcommand\aastexcls{2}
\newcommand\othercls{3}

\newcommand\papercls{\aastexcls}
\documentclass[tighten, times, twocolumn]{aastex61}  % ApJ look-alike
\if\papercls \apjcls
\usepackage{apjfonts}
\else\if\papercls \othercls
\usepackage{epsfig}
\fi\fi
\usepackage{ifthen}
\usepackage{natbib}
\usepackage{amssymb, amsmath}
\usepackage{appendix}
\usepackage{etoolbox}
\if\papercls \apjcls
\newcommand\aas{\ref@jnl{AAS Meeting Abstracts}}% *** added by jh
\newcommand\dps{\ref@jnl{AAS/DPS Meeting Abstracts}}% *** added by jh
\newcommand\maps{\ref@jnl{MAPS}}% *** added by jh
\else\if\papercls \othercls
\usepackage{astjnlabbrev-jh}
\fi\fi

\if\papercls \apjcls
\bibpunct[, ]{(}{)}{,}{a}{}{,}
\else
\bibpunct[, ]{(}{)}{,}{a}{}{,}
\fi

\if\papercls \aastexcls
\hypersetup{citecolor=blue, % color for \cite{...} links
            linkcolor=blue, % color for \ref{...} links
            menucolor=blue, % color for Acrobat menu buttons
            urlcolor=blue}  % color for \url{...} links
\else
\usepackage[%pdftex,      % hyper-references for pdflatex
bookmarks=true,           %%% generate bookmarks ...
bookmarksnumbered=true,   %%% ... with numbers
colorlinks=true,          % links are colored
citecolor=blue,           % color for \cite{...} links
linkcolor=blue,           % color for \ref{...} links
menucolor=blue,           % color for Acrobat menu buttons
urlcolor=blue,            % color for \url{...} links
linkbordercolor={0 0 1},  %%% blue frames around links
pdfborder={0 0 1},
frenchlinks=true]{hyperref}
\fi

\if\papercls \othercls
\newcommand{\eprint}[1]{\href{http://arxiv.org/abs/#1}{#1}}
\else
\renewcommand{\eprint}[1]{\href{http://arxiv.org/abs/#1}{#1}}
\fi

\providecommand{\adsurl}[1]{\href{#1}{ADS}}

\makeatletter
\patchcmd{\NAT@citex}
  {\@citea\NAT@hyper@{%
     \NAT@nmfmt{\NAT@nm}%
     \hyper@natlinkbreak{\NAT@aysep\NAT@spacechar}{\@citeb\@extra@b@citeb}%
     \NAT@date}}
  {\@citea\NAT@nmfmt{\NAT@nm}%
   \NAT@aysep\NAT@spacechar\NAT@hyper@{\NAT@date}}{}{}

\patchcmd{\NAT@citex}
  {\@citea\NAT@hyper@{%
     \NAT@nmfmt{\NAT@nm}%
     \hyper@natlinkbreak{\NAT@spacechar\NAT@@open\if*#1*\else#1\NAT@spacechar\fi}%
       {\@citeb\@extra@b@citeb}%
     \NAT@date}}
  {\@citea\NAT@nmfmt{\NAT@nm}%
   \NAT@spacechar\NAT@@open\if*#1*\else#1\NAT@spacechar\fi\NAT@hyper@{\NAT@date}}
  {}{}
\makeatother

\makeatletter
\DeclareRobustCommand{\lowcase}[1]{\@lowcase#1\@nil}
\def\@lowcase#1\@nil{\if\relax#1\relax\else\MakeLowercase{#1}\fi}
\makeatother

\DeclareSymbolFont{UPM}{U}{eur}{m}{n}
\DeclareMathSymbol{\umu}{0}{UPM}{"16}
\let\oldumu=\umu
\renewcommand\umu{\ifmmode\oldumu\else\math{\oldumu}\fi}
\newcommand\micro{\umu}
\if\papercls \othercls
\newcommand\micron{\micro m}
\else
\renewcommand\micron{\micro m}
\fi
\newcommand\microns{\micron}

\let\oldsim=\sim
\renewcommand\sim{\ifmmode\oldsim\else\math{\oldsim}\fi}
\let\oldpm=\pm
\renewcommand\pm{\ifmmode\oldpm\else\math{\oldpm}\fi}
\newcommand\by{\ifmmode\times\else\math{\times}\fi}
\newcommand\ttt[1]{10\sp{#1}}
\newcommand\tttt[1]{\by\ttt{#1}}

\newbox{\wdbox}
\renewcommand\c{\setbox\wdbox=\hbox{,}\hspace{\wd\wdbox}}
\renewcommand\i{\setbox\wdbox=\hbox{i}\hspace{\wd\wdbox}}

\newcount\timect
\newcount\hourct
\newcount\minct
\newcommand\now{\timect=\time \divide\timect by 60
         \hourct=\timect \multiply\hourct by 60
         \minct=\time \advance\minct by -\hourct
         \number\timect:\ifnum \minct < 10 0\fi\number\minct}

\catcode`@=11

\newcommand\comment[1]{}

\newcommand\commenton{\catcode`\%=14}

\renewcommand\math[1]{$#1$}
\newcommand\mathshifton{\catcode`\$=3}

\let\atab=&
\newcommand\atabon{\catcode`\&=4}

\let\oldmsp=\sp
\let\oldmsb=\sb
\def\sp#1{\ifmmode
           \oldmsp{#1}%
         \else\strut\raise.85ex\hbox{\scriptsize #1}\fi}
\def\sb#1{\ifmmode
           \oldmsb{#1}%
         \else\strut\raise-.54ex\hbox{\scriptsize #1}\fi}
\newbox\@sp
\newbox\@sb
\def\sbp#1#2{\ifmmode%
           \oldmsb{#1}\oldmsp{#2}%
         \else
           \setbox\@sb=\hbox{\sb{#1}}%
           \setbox\@sp=\hbox{\sp{#2}}%
           \rlap{\copy\@sb}\copy\@sp
           \ifdim \wd\@sb >\wd\@sp
             \hskip -\wd\@sp \hskip \wd\@sb
           \fi
        \fi}
\def\msp#1{\ifmmode
           \oldmsp{#1}
         \else \math{\oldmsp{#1}}\fi}
\def\msb#1{\ifmmode
           \oldmsb{#1}
         \else \math{\oldmsb{#1}}\fi}

\def\supon{\catcode`\^=7}

\def\subon{\catcode`\_=8}

\def\supsubon{\supon \subon}

\newcommand\actcharon{\catcode`\~=13}

\newcommand\paramon{\catcode`\#=6}

\comment{And now to turn us totally on and off...}

\newcommand\reservedcharson{ \commenton  \mathshifton  \atabon  \supsubon 
                             \actcharon  \paramon}

\catcode`@=12
\reservedcharson

\if\papercls \apjcls

\else

\fi

\if\papercls \othercls
\else
  \newcommand\inpress{n}
  \if\inpress y
    \received{\today}
    \revised{}
    \accepted{}
    \if\papercls \apjcls
    \slugcomment{}
    \fi
  \else
  \fi
\fi

\newcommand\chisq{\ifmmode{\chi\sp{2}}\else\math{\chi\sp{2}}\fi}
\newcommand\redchisq{\ifmmode{ \chi\sp{2}\sb{\rm red}}
                    \else\math{\chi\sp{2}\sb{\rm red}}\fi}

\newcommand\Teq{\ifmmode{T\sb{\rm eq}}\else\math{T\sb{\rm eq}}\fi}
\newcommand\mjup{$M\sb{\rm Jup}$}
\newcommand\rjup{$R\sb{\rm Jup}$}

\newcommand\der{\ifmmode{\rm d}\else\math{\rm d}\fi}
\newcommand\vs{\emph{vs.}}

\newcommand\molhyd{H$\sb{2}$}
\newcommand\methane{CH$\sb{4}$}
\newcommand\water{H$\sb{2}$O}
\newcommand\carbdiox{CO$\sb{2}$}

\shorttitle{Aerosol Constraints on WASP-49\lowcase{b}}
\shortauthors{Cubillos {\em et al.}}

\begin{document}

\title{Aerosol Constraints on the Atmosphere of the Hot Saturn-mass planet WASP-49\,\lowcase{b}}

\author{Patricio~E.~Cubillos}
\affiliation{Space Research Institute, Austrian Academy of Sciences,
             Schmiedlstrasse 6, A-8042, Graz, Austria}
\author{Luca~Fossati}
\affiliation{Space Research Institute, Austrian Academy of Sciences,
             Schmiedlstrasse 6, A-8042, Graz, Austria}
\author{Nikolai~V.~Erkaev}
\affiliation{Federal Research Center "Krasnoyarsk Science Center" SB RAS,
             "Institute of Computational Modelling", Krasnoyarsk 660036, Russia}
\author{Matej~Malik}
\affiliation{Center for Space and Habitability, University of Bern,
             Sidlerstrasse 5, CH-3012, Bern, Switzerland}
\author{Tetsuya~Tokano}
\affiliation{Institut f\"ur Geophysik und Meteorologie, Universit\"at
             zu K\"oln, Albertus-Magnus-Platz, 50923 K\"oln, Germany}
\author{Monika~Lendl}
\affiliation{Space Research Institute, Austrian Academy of Sciences,
             Schmiedlstrasse 6, A-8042, Graz, Austria}
\affiliation{Max Planck Institute for Astronomy,
             K\"onigstuhl 17, 69117 Heidelberg, Germany}
\author{Colin~P.~Johnstone}
\affiliation{Department of Astrophysics, University of Vienna,
             T{\"u}rkenschanzstrasse 17, 1180 Vienna, Austria}
\author{Helmut~Lammer}
\affiliation{Space Research Institute, Austrian Academy of Sciences,
             Schmiedlstrasse 6, A-8042, Graz, Austria}
\author{Aur\'elien Wyttenbach}
\affiliation{Geneva Observatory, University of Geneva,
             ch. de Maillettes 51, CH-1290 Versoix, Switzerland}

\email{patricio.cubillos@oeaw.ac.at}

\begin{abstract}
The strong, nearly wavelength-independent absorption cross section of
aerosols produces featureless exoplanet transmission spectra, limiting
our ability to characterize their atmospheres.  Here we show that even
in the presence of featureless spectra, we can still characterize
certain atmospheric properties.  Specifically, we constrain the upper
and lower pressure boundaries of aerosol layers, and present plausible
composition candidates.  We study the case of the bloated Saturn-mass
planet WASP-49\,b, where near-infrared observations reveal a flat
transmission spectrum between 0.7 and 1.0 {\microns}.  First, we use a
hydrodynamic upper-atmosphere code to estimate the pressure reached by
the ionizing stellar high-energy photons at $\ttt{-8}$ bar, setting
the upper pressure boundary where aerosols could exist.  Then, we
combine HELIOS and Pyrat Bay radiative-transfer models to
constrain the temperature and photospheric pressure of atmospheric
aerosols, in a Bayesian framework.
For WASP-49\,b, we constrain the transmission photosphere (hence, the
aerosol deck boundaries) to pressures above $\ttt{-5}$
bar \added{(100$\times$ solar metallicity), $\ttt{-4}$ bar (solar),
and $\ttt{-3}$ bar (0.1$\times$ solar) as lower boundary, and below
$\ttt{-7}$ bar as upper boundary.}
Lastly, we compare condensation curves of aerosol compounds with the
planet's pressure-temperature profile to identify plausible
condensates responsible for the absorption.  Under these
circumstances, we find as candidates: Na$\sb{2}$S (at 100$\times$ solar
metallicity); Cr and MnS (at solar and 0.1$\times$
solar); \added{and forsterite, enstatite, and alabandite (at 0.1$\times$
solar)}.
\end{abstract}

% http://journals.aas.org/authors/keywords2013.html
\keywords{
planets and satellites: atmospheres --
planets and satellites: individual: WASP-49\,b --
methods: numerical --
techniques: spectroscopic}

\section{INTRODUCTION}
\label{introduction}

Over the past decade, photometric observations of transiting
exoplanets have become the main tool to characterize exoplanet
atmospheres.  Transit or eclipse events provide a direct measurement
of the transmission or emission spectrum of a planetary atmosphere,
respectively.  The temperature and composition of an atmosphere
modulate an observed spectrum, as each atmospheric species has a very
specific spectral absorption pattern.  Therefore, multi-wavelength
observations allow us to disentangle the contribution of different
species, constraining atmospheric properties.

Unfortunately, characterizing atmospheres has been proven to be a more
challenging effort than expected, since many exoplanet observations
show nearly featureless spectra \citep[e.g.,][]{Pont2008HD189Haze,
MandellEtal2013apjWFC3specWASP12b-17b-19b, KnutsonEtal2014natGJ436b,
KreidbergEtal2014natCloudsGJ1214b}.  Muted spectral features are
attributed to the presence of cloud condensates and photochemical
hazes (hereafter, aerosols), whose strong opacities and weak
wavelength dependence obscure other spectral features from deeper
regions of an atmosphere.  In light of the ubiquity of aerosol
features in exoplanet atmospheres, researchers have adopted two
different approaches, either study the prominence of cloud-covered
atmospheres to avoid selecting these targets, or better characterize
atmospheric aerosol properties.

The current sample of well-studied exoplanet atmospheres (good
spectral coverage and data quality) is just enough to enable tentative
trends in the cloud prominence.
\citet{Stevenson2016apjlCloudProminence} found a higher cloud
prominence for the more temperate (equilibrium temperature $\Teq<700$
K) and lower surface gravity planets ($\log g<2.8$), based on the
strength of the 1.4 {\micron} {\water} band.
\citet{Heng2016apjlCloudinessIndex} found a
tentative lower cloudiness index for the more irradiated atmospheres,
based on the alkali absorption line profiles.
Lastly, \citet{BarstowEtal2017apjHotJupiterRetrieval} found evidence
of aerosol absorption in all 10 planets
from \citet{SingEtal2016natClearCloudyContinuum}.
Overall, finding that planets with $1300\ {\rm K} <\Teq <1700$ K, are
more consistent with gray cloud layers, whereas other planets are more
consistent with strong Rayleigh scattering absorption.

Although the lack of features of aerosols makes them intrinsically
hard to characterize, several studies have improved our understanding
of their properties and consequences on exoplanet atmospheres and
spectra.  For example,
\citet{WakefordSing2015aaHotJupiterClouds}
studied the effects of grain sizes and distributions on hot-Jupiter
transmission spectra, finding absorption features which could
differentiate condensate formation scenarios, such as condensate
clouds or photochemically generated species.
\citet{MorleyEtal2015apjFlatModelSpectra} determined variations
in the optical albedo between cloudy (moderate) and hazy (dark)
atmospheres for warm planets, or thermal inversions caused by hazes.
\citet{ParmentierEtal2016apjHotJupiterClouds} have studied how
variations in the cloud composition with equilibrium temperature shape
the transmission spectrum of hot Jupiter atmospheres.
These and other theoretical studies help us to provide more accurate
diagnostics when we encounter cloud-dominated observations.

In this article we further explore on the characterization of cloudy
atmospheres, investigating what properties can we constrain given
present-day ground-based transmission spectra.  We study the case of
the bloated Saturn-mass but Jupiter-sized planet WASP-49\,b, which
presents a flat near-infrared (NIR) transmission spectrum.  By applying a mixed
forward and retrieval-modeling approach, we characterize the
atmosphere of WASP-49\,b with a combination of hydrodynamic,
radiative-transfer, and equilibrium-condensation atmospheric models.
Assuming that the observations are the result of an optically thick
aerosol layer, we constrain the pressure boundaries of the layer, and
then list plausible aerosol condensates for different atmospheric
metallicity scenarios.  In Section \ref{sec:obs}, we summarize the
properties and observations of the WASP-49 system.  In
Section \ref{sec:analysis}, we model the atmosphere of
WASP-49\,b.  Finally, in Section \ref{sec:conclusions} we present our
conclusions.

\section{WASP-49\,\lowercase{b} TRANSMISSION SPECTRUM}
\label{sec:obs}

\begin{figure}[t]
\centering
\includegraphics[width=\linewidth, clip]{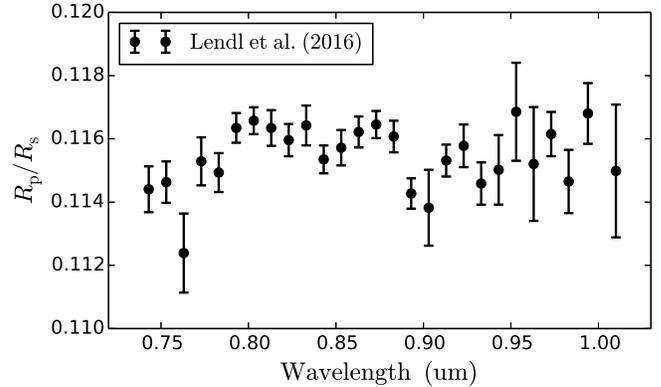}
\caption{
WASP-49\,b FORS2 transmission spectrum
from \citet{LendlEtal2016aaWASP49bTransmissionFORS2}. For a clear
atmosphere, K and {\water} absorption features should increase
the modulation spectrum at 0.78 {\microns} and 0.95 {\microns},
respectively.}
\label{fig:transmission}
\end{figure}

The hot Jupiter-sized exoplanet
WASP-49\,b \citep{LendlEtal2012aaWASP49b} has a mass of $M\sb{\rm
p}=0.40 \pm 0.03$ {\mjup}, radius $R\sb{\rm p}=1.2\pm0.05$ {\rjup},
and equilibrium temperature $\Teq = 1400 \pm 40$~K (assuming
negligible Bond
albedo).  \citet{LendlEtal2016aaWASP49bTransmissionFORS2} reported
multiple NIR broad-band transmission observations with the FORS2
instrument of the ESO/VLT, from 0.7
{\microns} to 1.0 {\microns}.  These observations revealed a nearly
flat, featureless optical spectrum (Figure \ref{fig:transmission}).
The lack of spectral features in the transmission spectra suggests
that the atmosphere of WASP-49\,b has an optically thick cloud deck,
blocking the atomic and molecular features.

For a clear atmosphere, the NIR spectrum is particularly useful to
constrain the atmospheric composition, as there are strong and
isolated molecular and atomic features.  Atmospheric sodium and
potassium produce strong, broad, and localized absorption lines
(centered at 0.59 and 0.77~{\microns}, respectively).  {\water} is the
most abundant molecule spectroscopically active in this range.  At
solar abundances, {\water} produces a clear absorption band at
0.9--1.0~{\microns}.  If present, TiO and VO dominate most of this
region of the spectrum.  However, the heavy TiO and VO molecules can
easily condense and rain out toward higher pressures for
temperatures lower than $\sim$1500 K \citep[][]{FortneyEtal2008apjTwoClasses,
SpiegelEtal2009apjTiO}.  We do not expect to find TiO and VO in the
atmosphere of WASP-49\,b at the altitude sampled by the FORS2
observations.

\section{ATMOSPHERIC MODELING}
\label{sec:analysis}

To constrain the aerosol properties of WASP-49\,b, we model the
planetary atmosphere with a succession of models.  First, we study the
upper atmosphere of the planet with hydrodynamic models to determine
the depth of the stellar high-energy irradiation, constraining the
upper boundary of the aerosol layer.  Then, we use
radiative-transfer models to retrieve the pressure
corresponding to the transit radius of the planet (the photospheric
pressure).  This is the range of pressures where the aerosols
condensate, making the atmosphere optically thick.  Finally, we use
the constraint on the aerosol pressure boundaries to propose plausible
condensate species responsible for the flat transmission spectra, by
comparing the planet temperature profile to condensation curves of
known condensates.

\subsection{Hydrodynamic Modeling}
\label{sec:hydro}

The stellar incident high-energy irradiation drives the planetary
upper-atmosphere chemistry and dynamics.  In particular, stellar X-ray
and extreme-ultraviolet (XUV) photons \added{deposit large amounts of
energy into the atmosphere, raising the temperature and} dissociating
atmospheric molecules.  The effective XUV radius (where most of the
stellar XUV flux is deposited) sets the upper boundary where aerosols
can exist.

To study the upper atmosphere of WASP-49\,b, we apply the 1D
hydrodynamic model of \citet{ErkaevEtal2016mnrasThermalLoss}.  This
model solves the mass, momentum, and energy-conservation system of
equations, allowing us to derive the pressure, temperature, and
composition profiles of WASP-49\,b, fully considering the stellar XUV
irradiation, Ly-$\alpha$ cooling, and atmospheric ionization,
dissociation, and recombination.  This model considers a simplified
hydrogen chemistry, and thus we consider the result from this model
only for upper layers of an atmosphere, where most of the atmospheric
particles are
dissociated \citep{KoskinenEtal2013icarPhotodynamicModel}.

We estimate the XUV luminosities for WASP-49\,b using the scaling laws
of \citet{WrightEtal2011apjStellarActivityRotation} to convert the
stellar rotation rates and masses into X-ray luminosities,
and \citet{SanzForcadaEtal2011aaXUVevaporation} to convert the X-ray
luminosities into extreme-ultraviolet luminosities.  At the orbit of
the planet, we obtain X-ray and extreme-ultraviolet fluxes of 240
erg\,s$\sp{-1}$cm$\sp{-2}$ and 2500 erg\,s$\sp{-1}$cm$\sp{-2}$,
respectively.

\begin{figure}[t]
\centering
\includegraphics[width=\linewidth, clip]{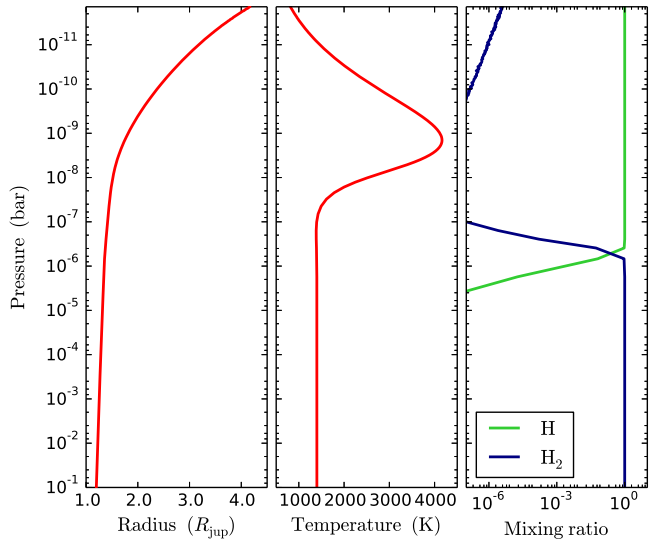}
\caption{
WASP-49\,b atmospheric profile derived from the hydrodynamic model. The
left, middle, and right panels show the radius, temperature, and
composition profiles for WASP-49\,b as function of pressure.}
\label{fig:hydroruns}
\end{figure}

Figure \ref{fig:hydroruns} shows the output atmospheric profiles from
the hydrodynamic model for WASP-49\,b.  The abrupt increase in
temperature at \sim$\ttt{-8}$ bar reveals where most of the stellar
high-energy irradiation is being deposited, and therefore, marks the
XUV effective radius.  Hydrogen dissociates at slightly higher
pressures, between $\ttt{-6}$ bar and $\ttt{-7}$ bar.  \added{The
extremely high temperatures reached at the XUV effective radius
(\sim4000~K) are sufficient to evaporate any atmospheric
condensate. Therefore, we set the $\ttt{-8}$ bar level} as the
uppermost boundary where aerosol condensates can remain.  The XUV
effective radius remains approximately at the same pressure when we
change the stellar irradiation fluxes by an order of magnitude.

\added{
We estimate the particle falling timescales to investigate its role in
determining the presence of cloud particles at these low pressures.
The falling timescale of particles is $\tau_f=H/v_t$, where $H$ is the
pressure scale height and $v_t$ is the terminal velocity of
particles.  The terminal velocity for each condensed species can be
calculated \citep[e.g., ][]{SpiegelEtal2009apjTiO} as
\begin{equation}
v_t=\frac{2}{9}\frac{a^2 \rho_c g}{\eta},
\end{equation}
where $a$ is the particle radius, $\rho_c$ is the mass density of the
condensate, $g$ is the gravitational acceleration and $\eta$ is the
dynamic viscosity of air.

The mass density of the condensates considered in this study lies in
the range between 1170 $\rm kg\ m^{-3}$ (for $\rm NH_4SH$) and 7874
$\rm kg\ m^{-3}$ (for Fe). The gravitational acceleration of WASP-49b
amounts to 7.18 $\rm m\ s^{-2}$ with the estimated mass and radius of
this planet. The dynamic viscosity of gaseous $\rm H_2$ at 1000 K is
0.019 Pa s according to the gas viscosity
calculator\footnote{\href{https://www.lmnoeng.com/Flow/GasViscosity.php}
{https://www.lmnoeng.com/Flow/GasViscosity.php}}.  The scale height of
an $\rm H_2$ atmosphere at 1000 K with the above $g$ amounts to
$5.75\times 10^5$ m.

These numbers yield $v_t$ of 9.8--$66.1\times 10^{-8}\ \rm m\ s^{-1}$
for a 1 $\mu$m-sized cloud particle and 9.8--$66.1 \times 10^{-4}\rm
m\ s^{-1}$ for a 100 {\micron}-sized particle.  Consequently, the
falling timescale is 0.9--$5.9\times 10^{12}$ s for 1 {\micron} and
0.9--$5.9\times 10^{8}$ s for 100 {\micron}.  Thus, the resulting
terminal velocities are small, as is typical for cloud particles.}

\subsection{Radiative-transfer Modeling}
\label{sec:RT}

The flat broad-band transmission spectrum of WASP-49\,b not only
suggests that the planet has an aerosol layer, it also constrains the
altitude where the atmosphere becomes opaque, i.e., the photosphere.
However, the corresponding photospheric pressure remains a degenerate
parameter that depends on the atmospheric temperature, composition,
and aerosol opacity.  Here, we combine forward and retrieval
radiative-transfer models to constrain the photospheric pressure of
WASP-49\,b.

\subsubsection{Radiative-equilibrium Temperature Model}
\label{sec:radeq}

To get a first idea of the atmospheric state of WASP-49\,b, we compute
radiative-equilibrium atmospheric models using the HELIOS
radiative-transfer code \citep{MalikEtal2017apjHelios}.  The goal is
to obtain representative atmospheric temperature values by running
ad-hoc simulations with and without aerosol layers.  The
radiative-transfer runs consider \added{infrared and
shorter-wavelength} opacities from {\water}, {\carbdiox}, and
CO \added{\citep{RothmanEtal2010jqsrtHITEMP}};
{\methane}, \added{NH$_3$, HCN, and
C$_2$H$_2$ \citep{RothmanEtal2013jqsrtHITRAN};
H$_2$S \citep{AzzamEtal2016mnrasExomolH2S}; the alkali metals Na and
K \citep{HengEtal2015apjSodiumLines, Heng2016apjlCloudinessIndex}};
collision-induced absorption from {\molhyd}-{\molhyd} and
{\molhyd}-He \added{\citep{RichardEtal2012jqsrtCIA}}; and {\molhyd}
Rayleigh scattering.

The atmospheric model consists of a one-dimensional
log-spaced-pressure model, extending from $\ttt{3}$ bar to $\ttt{-9}$
bar.  We compute radiative-equilibrium temperature profiles subject to
an emission brightness temperature of 1400~K, i.e., the equilibrium
temperature WASP-49\,b.  The atmosphere receives an incident stellar
irradiation modeled with a PHOENIX spectrum, according to the
parameters of \citet{LendlEtal2016aaWASP49bTransmissionFORS2}.  We
keep a fixed solar-metallicity composition in thermochemical
equilibrium \added{(Stock et al. 2017, in prep.).  The chemistry model
includes around 550 gas-phase species and is based on the
semi-analytic approach described
in \citet{GailSedlmayr2013PhysChemDust}.
}

\begin{figure}[t]
\centering
\includegraphics[width=\linewidth, clip]{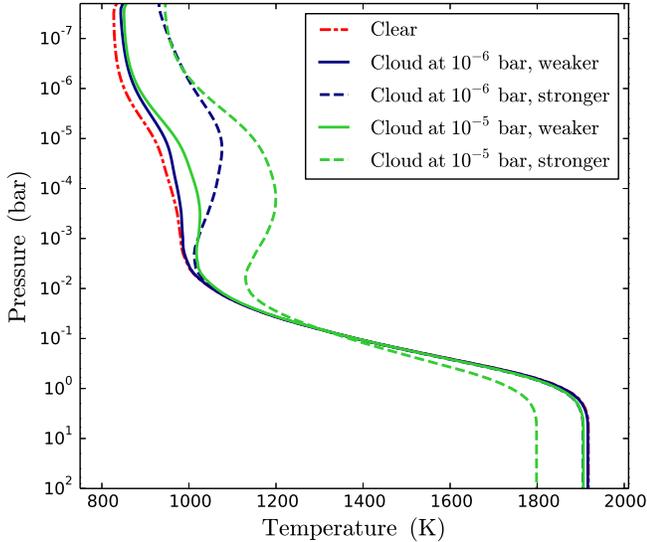}
\caption{
HELIOS radiative-equilibrium temperature profiles for WASP-49\,b.  The
models are constrained such that the the total output flux corresponds
to a brightness temperature of 1400~K.  All models consider the main
opacity sources relevant for Jupiter-like exoplanets.  The cloud
models consist of a log-normally distributed opacity in pressure with
total opacity $k = \ttt{-2}$ cm$\sp{2}$g$\sp{-1}$ (weaker) and $\ttt{-1}$
cm$\sp{2}$g$\sp{-1}$ (stronger, see legend).}
\label{fig:helios}
\end{figure}

To simulate an aerosol layer, we include a log-normally distributed
gray-opacity absorber over pressure.  We set the standard deviation of
the distribution to 1.0 in $\log\sb{10}$(pressure) and test two
altitudes for the peak, $\ttt{-5}$~bar and $\ttt{-6}$~bar.  We also
model two cases for the total aerosol opacity
of $k = \ttt{-2}$~cm$\sp{2}$g$\sp{-1}$ and
$\ttt{-1}$~cm$\sp{2}$g$\sp{-1}$.
\added{These clouds have non-isotropic scattering and a non-vanishing
  albedo.  We estimated these parameters from MgSiO$_3$ aerosol
  particles with size $\sim$5 {\micron}, using a Mie-scattering code.
  Since we do not know the cloud composition, this condensate provides
  a good first-order guess for silicate albedos.  The obtained albedos
  range on the order of a few percent, with the lower
  clouds the maximum albedo, up to $A\approx 0.17$.  As silicate clouds are
  mostly forward scattering (asymmetry parameter $g_0\sim0.7$--0.9),
  their absorption effect outweighs their reflective effect of stellar
  radiation.  Thus, the net effect of the clouds causes a global
  warming of atmospheric temperatures.
}

Figure \ref{fig:helios} shows the radiative-equilibrium temperature
profiles for WASP-49\,b.  \added{The greenhouse gasses, namely
{\water}, {\carbdiox}, and {\methane}, dominate the absorption at
depth (and thus, the heating).} The aerosol layer produces a
weak-to-moderate temperature increase with respect to the
clear-atmosphere model.  The peak of the heating is located a few
scale heights below the aerosol layer, and is proportional to the
aerosol opacity, as expected.  \added{The Na and K resonance doublet
lines dominate the absorption at the highest layers of the atmosphere
(above \sim$10^{-7}$--$10^{-8}$ bar).  However, at these high
altitudes, the atmosphere is already directly interacting with the
high-energy stellar irradiation, where the radiative-equilibrium
assumptions (i.e., hydrostatic and thermo-chemical equilibrium) do not
hold anymore.  Thus, we trust the radiative-equilibrium profiles up to
the $\sim\ttt{-8}$ bar level.  Between pressures of 0.1 and $\ttt{-8}$
bar (the region probed by NIR transmission spectroscopy),} the
atmospheric temperature ranges between 800 K and 1200 K.

\subsubsection{Transmission Photospheric-Pressure Retrieval}
\label{sec:retrieval}

\begin{figure}[t]
\centering
\includegraphics[width=\linewidth, clip]{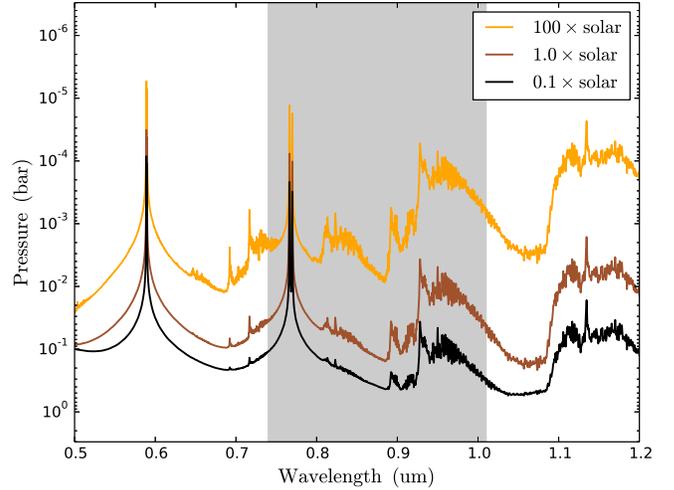}
\caption{WASP-49\,b clear-atmosphere model transmission spectra.
This is the pressure corresponding to the transit radius modulation.
The different models have scaled solar metallicities (see legend),
with abundances in thermochemical equilibrium at the
planet's equilibrium temperature (1400 K).  The shaded area denotes the
wavelength range probed by the FORS2 transmission observations.
Aerosol layers would need to sit above these pressures to hide these
spectral features.}
\label{fig:clear}
\end{figure}

The observed NIR transmission spectrum of WASP-49\,b
(Fig.\ \ref{fig:transmission}) constrains the transit radius of the
planet.  However, this dataset does not directly constrain the
 pressure ($p\sb{\rm T}$) at the transit
radius.  Assuming that the flat NIR spectrum of WASP-49\,b is the result
of an opaque aerosol layer, upon additional considerations, we can
obtain posterior distributions for $p\sb{\rm T}$, and hence, the cloud
top level.  An opaque aerosol layer must lay high enough in the
atmosphere to blanket other spectral features remain unobserved; in
this case, the {\water} band at 0.95 {\micron}, the K doublet at 0.77
{\microns}, and the Rayleigh {\molhyd} slope (Figure \ref{fig:clear}).
Finding this pressure is a degenerate problem that hinges also on the
unconstrained atmospheric temperature and composition, beside the
aerosol properties themselves.

Everything considered, we adopt a Bayesian retrieval approach to
constrain the transmission photospheric pressure of the planet.  To
this end, we model the NIR transmission spectrum of WASP-49\,b with the
Python Radiative-transfer in a Bayesian framework package (Pyrat
Bay, \citeauthor{CubillosEtal2017apjPyratBay}, in prep.).  Pyrat Bay
is a state-of-the-art, open-source, reproducible
package\footnote{\href{https://github.com/pcubillos/pyratbay}
{https://github.com/pcubillos/pyratbay}.}, which is an update of the
Bayesian Atmospheric Radiative Transfer
package \citep{Cubillos2016phdThesis, Blecic2016phdThesis}.  To
explore the parameter space, Pyrat Bay implements the
differential-evolution Markov-chain Monte Carlo (MCMC)
algorithm \citep[][]{CubillosEtal2017apjRednoise}.

The planet's atmospheric model consist of a one-dimensio-nal set of
spherically concentric shells (layers), equidistantly spaced in log
pressure.  We considered 100 layers ranging from $\ttt{-8}$ bar to 100
bar.  For transmission geometry, Pyrat Bay computes the fraction of
absorbed stellar flux for parallel rays (along the star--planet--Earth
line of sight) crossing the planetary atmosphere.

To explore different composition scenarios, we test three cases with
0.1, 1.0, and 100 times enhanced solar elemental metallicities.  For
each case we compute thermochemical-equilibrium
abundances \citep{BlecicEtal2016apsjTEA}, which we keep fixed during
the MCMC exploration.  Our atmospheric models include all major
species expected in gas-giant planets ({\water}, {\carbdiox}, CO,
{\methane} mainly).  However, only {\water} and K produce strong
enough absorption features in the observed spectral range.  We compute
the {\water} line-transition opacity from the HITEMP
database \citep[][]{RothmanEtal2010jqsrtHITEMP}.
We also include collision-induced absorption opacities for
{\molhyd}--{\molhyd} \citep{Borysow2002jqsrtH2H2lowT,
BorysowEtal2001jqsrtH2H2highT} and
{\molhyd}--He \citep{BorysowEtal1988apjH2HeRT,
BorysowEtal1989apjH2HeRVRT, BorysowFrommhold1989apjH2HeOvertones};
resonant Na and K lines \citep{BurrowsEtal2000apjBDspectra}; and
Rayleigh scattering
opacity \citep{LecavelierDesEtangsEtal2008aaRayleighHD189}.  We do not
expect to find TiO in the atmosphere since TiO condenses around
2000--1400~K for pressures between 1 bar and $\ttt{-8}$
bar \citep{SpiegelEtal2009apjTiO}.

There is a large variety of candidate aerosol condensates.  Each one
forms and remains in the atmosphere at specific pressure and
temperature conditions, but overall, covering a broad range of the
parameter
space \citep[e.g.,][]{WakefordSing2015aaHotJupiterClouds}.
Unfortunately, \added{when considering a limited wavelength region,}
most condensates produce similar, nearly featureless spectra, making
it hard to spectroscopically distinguish one from another.  Thus, it
is impractical to determine the specific condensates in a retrieval.
Instead, we model and fit a generic semi-infinite gray absorber with
constant opacity cross-section.

Given the penetration depth of the stellar high-energy flux estimated
by our upper-atmosphere hydrodynamic models (Section \ref{sec:hydro}),
we set the aerosol top-boundary pressure at $\ttt{-8}$ bar.  We leave the
aerosol opacity cross-section ($f\sb{\rm gray} \by\sigma\sb{0}$) as an
 MCMC free parameter, with $\sigma\sb{0}=5.3\tttt{-27}$
cm$\sp{2}$\,molec$\sp{-1}$ (the {\molhyd} opacity cross-section at 0.7
{\microns}) and $f\sb{\rm gray}$ a dimensionless scaling factor.  The
aerosol extinction coefficient (in cm$\sp{-1}$) at each layer results
from multiplying the cross-section with the {\molhyd} number density.
Therefore, the aerosol extinction coefficient scales linearly with
pressure.  Note that we use the {\molhyd} just as reference values for
the parameters.

We assume hydrostatic equilibrium to relate the pressure and radius of
the atmospheric layers.  To obtain the particular solution to the
hydrostatic-equilibrium differential equation we need to include a
`boundary' condition $p(r\sb{0}) = p\sb{0}$, where $p$ is the atmospheric
pressure and $p\sb{0}$ the pressure at the radius $r\sb{0}$.  Here, we
fix $p\sb{0}$ at 0.1 bar,
and leave $r\sb{0} \equiv R\sb{0.1\ {\rm bar}}$ as an MCMC free
parameter.  The $p\sb{0}$ and $r\sb{0}$ pair is implicitly constrained
by the transmission spectrum, although it typically degenerates with
other atmospheric parameters.

\begin{figure}[t]
\centering
\includegraphics[width=\linewidth, clip]{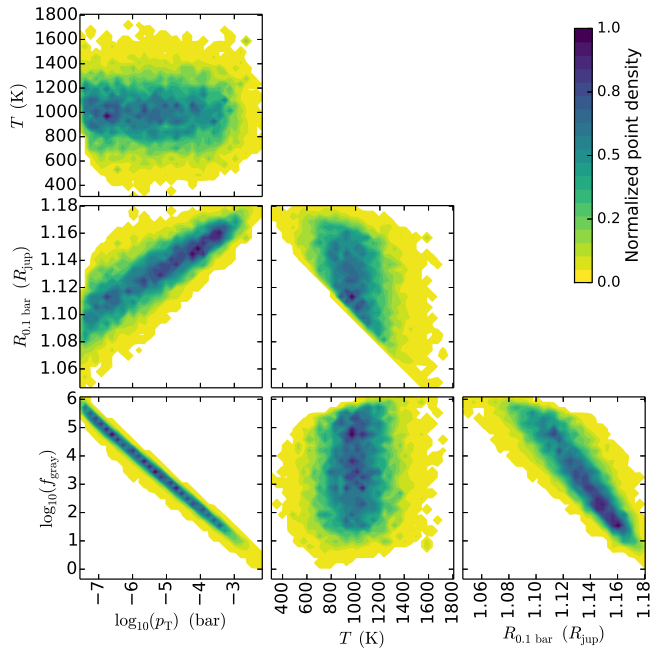}
\caption{MCMC pairwise posterior distribution for the solar-abundance
run.  Note that only $T$, $R\sb{0.1\ {\rm bar}}$, and $f\sb{\rm gray}$
are the MCMC free parameters, the photospheric pressure $p\sb{\rm T}$
is a derived parameter (see text).}
\label{fig:solarpair}
\end{figure}

The unknown temperature profile influences the output transmission
spectrum by modifying both the line-transition opacity of the species
and the hydrostatic-equilibrium solution.  The degeneracy with other
atmospheric properties renders the temperature mostly unconstrained by
the MCMC, given our limited transmission data.  Thus, our model adopts
a simple isothermal profile, with the temperature $T$ an MCMC free
parameter.  This is an appropriate model, considering that the
radiative-equilibrium models show temperature profiles that do not
dramatically change above 0.1 bar---the temperature at deeper layers
are irrelevant since those pressures are not accessible by the
transmission observation (see Fig.\ \ref{fig:clear}).  We further
limit the temperature range explored by the MCMC by including a
Gaussian prior centered at $1200$~K and with standard deviation of
$200$~K.  This range covers the temperatures sampled by the
radiative-equilibrium models.  This prior effectively prevents the
MCMC to sample extremely low temperatures which, however physically
plausible, are unlikely.  Later, the posterior distributions will show
that this prior does not influence the conclusions of this
work.

\begin{figure}[t]
\centering
\includegraphics[width=\linewidth, clip]{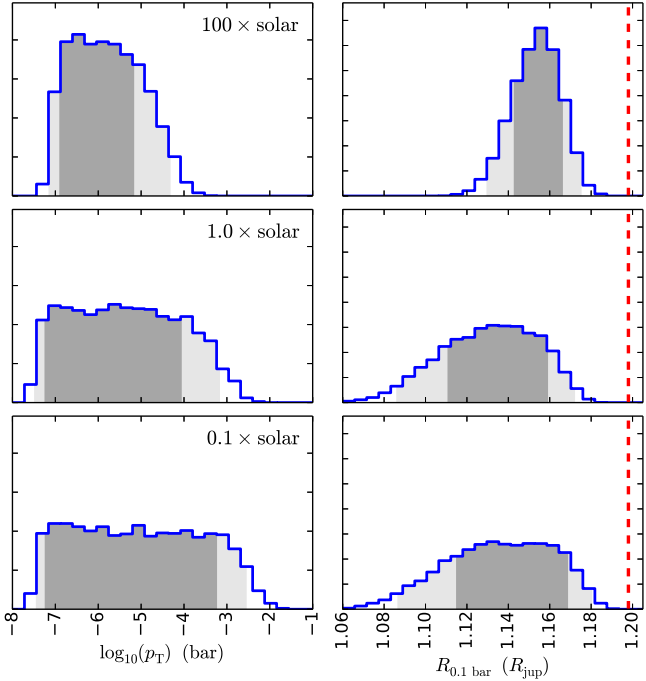}
\caption{Marginal posterior density of $p\sb{\rm T}$ and
$R\sb{0.1\ {\rm bar}}$ {\vs} metallicity.  The histograms are plotted
to scale such that the area under the curve is one.  The light and
dark gray areas denote the 95\% and 68\% highest-posterior-density (HPD)
credible regions, respectively.  The red dashed vertical line denotes the
\added{NIR} transit radius of WASP-49\,b, observed \added{by FORS2}.}
\label{fig:marginalZ}
\end{figure}

\subsubsection{WASP-49\,b Retrieval Results}
\label{sec:w49results}

For WASP-49\,b, we ran an MCMC for each of the three composition
scenarios, with $T$, $R\sb{0.1\ {\rm bar}}$, and $\log(f\sb{\rm
gray})$ the retrieval MCMC free parameters.
Figure \ref{fig:solarpair} shows the pairwise MCMC posterior
distributions for the solar-abundance run.  The runs for 0.1 and 100
times solar metallicity are qualitatively similar.
Additionally, knowing the pressure and radius profile for each
iteration, we derive the pressure corresponding to the observed
transit radius, $p\sb{\rm T} = p(R\sb{p})$.  As seen in
Figure \ref{fig:solarpair}, the main factor that
determines $p\sb{\rm T}$ is the aerosol opacity, and thus, this
transit pressure effectively constrains the top of the aerosol layer.

The correlation between free parameters in the posteriors reflects the
degeneracy of the fit.  Among these parameters, the temperature is the
least constrained by the data.  Consequently, the marginal posterior
distribution for the temperature replicates the adopted prior
distribution.
Although the arbitrariness in the temperature prior selection
expresses in the MCMC posteriors, this has a minor effect on the
photospheric pressure.  While
there are correlations between $p\sb{\rm T}$ and $R\sb{0.1\ {\rm
bar}}$ and $\log(f\sb{\rm gray})$, there is weak to no correlation
between $p\sb{\rm T}$ and $T$.

\begin{figure*}[t]
\centering
\includegraphics[width=1.0\linewidth, clip]{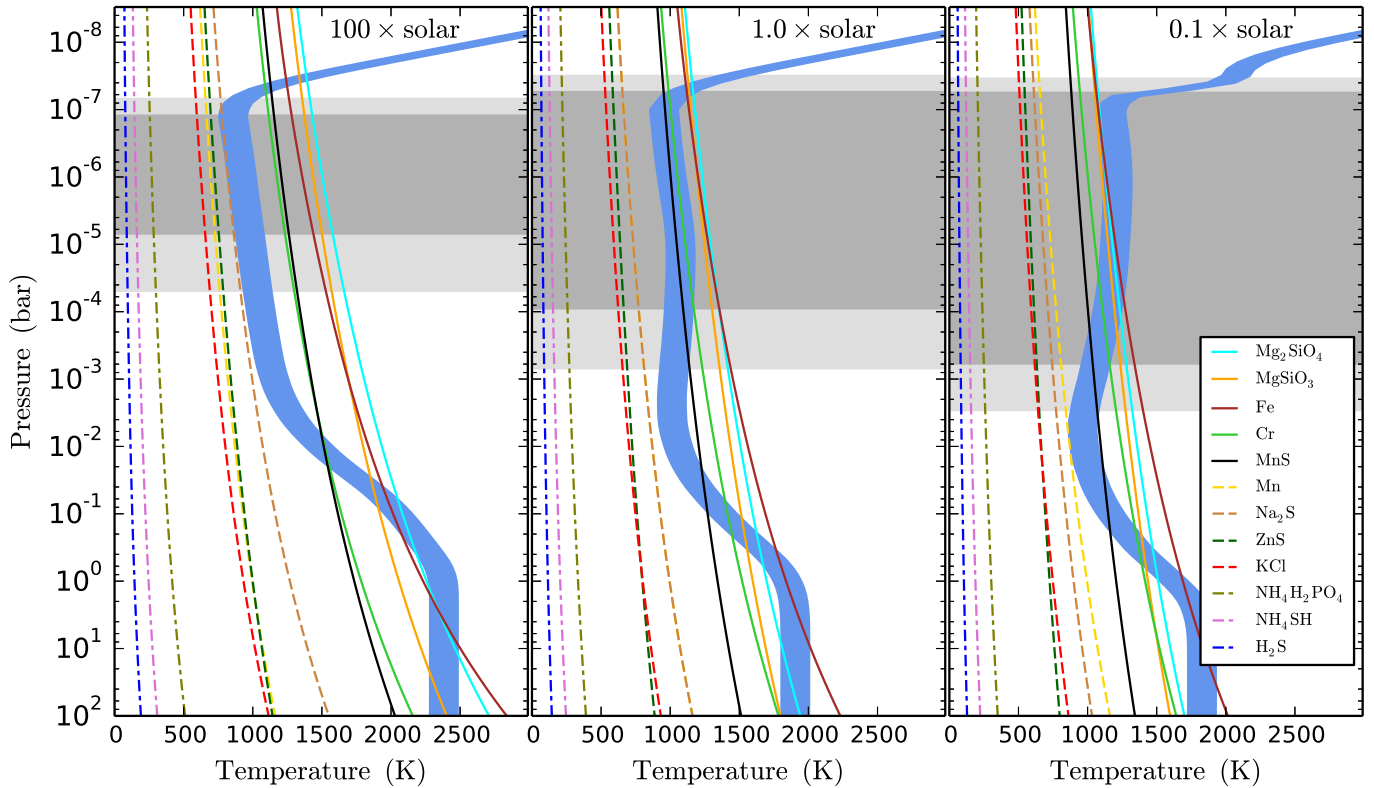}
\caption{Equilibrium condensation curves for a set of minerals 
as a function of pressure (see legend).  Each of the three panels
corresponds to an atmospheric metallicity case.  The light and dark
gray areas denote the 95\% and 68\% HPD credible regions
(respectively) of the photospheric pressure
(Section \ref{sec:retrieval}).  The blue band is the
radiative-equilibrium temperature profile (Section \ref{sec:radeq})
for the cloudy case at $\ttt{-6}$ bar with stronger opacity.  \added{The top of the temperature profile follows the XUV irradiation heat-up curve determined from the hydrodynamic models.} The
temperature-profile curve has a width of 200~K to consider a range of
possible temperatures. \comment{\vspace{0.5cm}}}
\label{fig:eqclouds}
\end{figure*}

Figure \ref{fig:marginalZ} shows the MCMC marginal posterior
distributions for $p\sb{\rm T}$ and $R\sb{0.1\ {\rm bar}}$, for each
of the composition case.
Depending on the atmospheric metallicity,
the atmospheric-retrieval runs indicate that an aerosol layer
should locate between \sim$\ttt{-7}$~bar (at the top) and
\sim$\ttt{-3}$~bar to \sim$\ttt{-5}$~bar (at the bottom)
to explain the flat transmission spectrum (68\%
highest-posterior-density, HPD).

The atmospheric metallicity plays an important role in constraining
the transmission photospheric pressure.
Given that the spectral features arise at lower pressures for a
higher-metallicity atmosphere (Fig.~\ref{fig:clear}), and that the
aerosols have to make the atmosphere optically thick at even lower
pressures (such that the model fit the flat transmission data), the
high-pressure boundary of $p\sb{\rm T}$ is located at lower pressures
for higher metallicities.  Furthermore, having no further
observational constraints, the lower-pressure boundary of $p\sb{\rm
T}$ is set by the XUV effective radius at $\ttt{-8}$ bar
(Section \ref{sec:hydro}).

\subsection{Equilibrium-clouds Modeling}

Now that we have constrained the pressure range where aerosols make
the atmosphere optically thick, we investigate which compounds can be
responsible for the flat transmission spectrum by comparing the
atmospheric temperature and pressure against equilibrium condensation
curves.

Our analysis is based on thermochemical equilibrium calculations that
are dominant at temperatures higher than
1000~K \citep{VisscherEtal2006apjAtmChemistryII,
VisscherEtal2010apjAtmChemistryIII,
MorleyEtal2012apjTYdwarfClouds}.
We assume that the atmosphere can contain an unspecified number of
mineral molecules in chemical equilibrium with these atmospheric
profiles.
Assuming elemental abundances in the solar
system \citep{Lodders2003apjCondensationTemperatures}, thermochemical
equilibrium determines the abundance of each species (partial
pressure) in equilibrium with a given temperature.  The partial pressure
can then be put into relation with the temperature-dependent
saturation vapor pressure of each species.  Once the partial pressure
of the species is determined, it can be converted to the equilibrium
condensation temperature, which can readily be compared to the
atmospheric temperature.  The equilibrium condensation temperature
generally increases with increasing total pressure and metallicity.
Deviations from thermochemical equilibrium can arise due to
photochemical reactions in the upper atmosphere or strong convective
mixing, but they are neglected in this approach.

Figure \ref{fig:eqclouds} depicts the equilibrium condensation
temperature profiles of 12 species for the three metallicities tested
(100, 1.0, and 0.1 times solar) of the atmosphere, calculated
after \citet{VisscherEtal2006apjAtmChemistryII,
VisscherEtal2010apjAtmChemistryIII}
and \citet{MorleyEtal2012apjTYdwarfClouds} depending on the species.
The equilibrium condensation temperature is the temperature below
which a given species can exist in condensed form under a given
pressure.

Having no observational temperature constrain (e.g., from a
secondary-eclipse observation), we consider HELIOS
radiative-equilibrium temperature models for each metallicity and with
a cloud layer centered at $\ttt{-6}$ bar and opacity of $k=\ttt{-1}
{\rm cm}\sp{2}{\rm g}\sp{-1}$.
\added{Varying the abundances has important effects on the infrared
cooling/heating of the atmosphere.  The higher abundance of greenhouse
gasses in the 100$\times$ solar model produces substantially warmer
temperatures at depth.}  However, \added{by definition,} the net
effective atmospheric temperature remains the same, as the photosphere
is simply pushed upward to cooler layers.  Additionally, the cloud
effect becomes diminished relative to the increased gaseous opacity,
as the latter increases with higher metallicity \added{(note that this
is a direct outcome of the modeling choice, as the aerosol opacity is
constant for each opacity)}.  If strong shortwave absorbers like TiO,
VO, \added{or a dark photochemical haze} were able to remain high in
the atmosphere, the increase in metallicity may lead to a temperature
inversion.

Given an atmospheric temperature profile, we can constrain which
condensates are consistent with the aerosol layer for each
metallicity.  Fig.~\ref{fig:eqclouds} shows that at 100 times solar
metallicity, sodium sulfide (Na$\sb2$S) is the most plausible
condensate in the atmosphere of WASP-49\,b.  As we explore lower
metallicities, the condensation curves shift toward lower
temperatures, and thus we expect alabandite (MnS) and Cr to intersect
the temperature profile at the pressures where we expect the aerosol
layer.  For sub-solar metallicities forsterite (Mg$\sb2$SiO$\sb4$),
enstatite (MgSiO$\sb3$), Fe, Cr, and alabandite are expected to
condense at the required pressures.  However, we note that, in
general, gas-giant planets are not expected to have sub-solar
metallicities \citep{KreidbergEtal2014apjWASP43b}.

\added{A caveat for this scenario is that some the condensates may rain out
of the upper atmospheres due to a ``cold trap'' phenomenon.
Since the temperature profiles cross some of the condensation curves
at more than one altitude (particularly the solar and sub-solar
cases), the condensed species are expected to be confined at the
deeper condensation point, \citep{HubenyEtal2003apjAtmospheres,
FortneyEtal2008apjTwoClasses}.  This would deplete the upper
atmosphere from the condensates, necessary to create the flat
transmission spectra.
\citet{SpiegelEtal2009apjTiO} invokes vigorous turbulent mixing on a 
macroscopic scale as a way to stir condensates up into
lower-pressure layers, even in the presence of a cold trap.
}

\begin{figure*}[t]
\centering
\includegraphics[width=1.0\linewidth, clip]{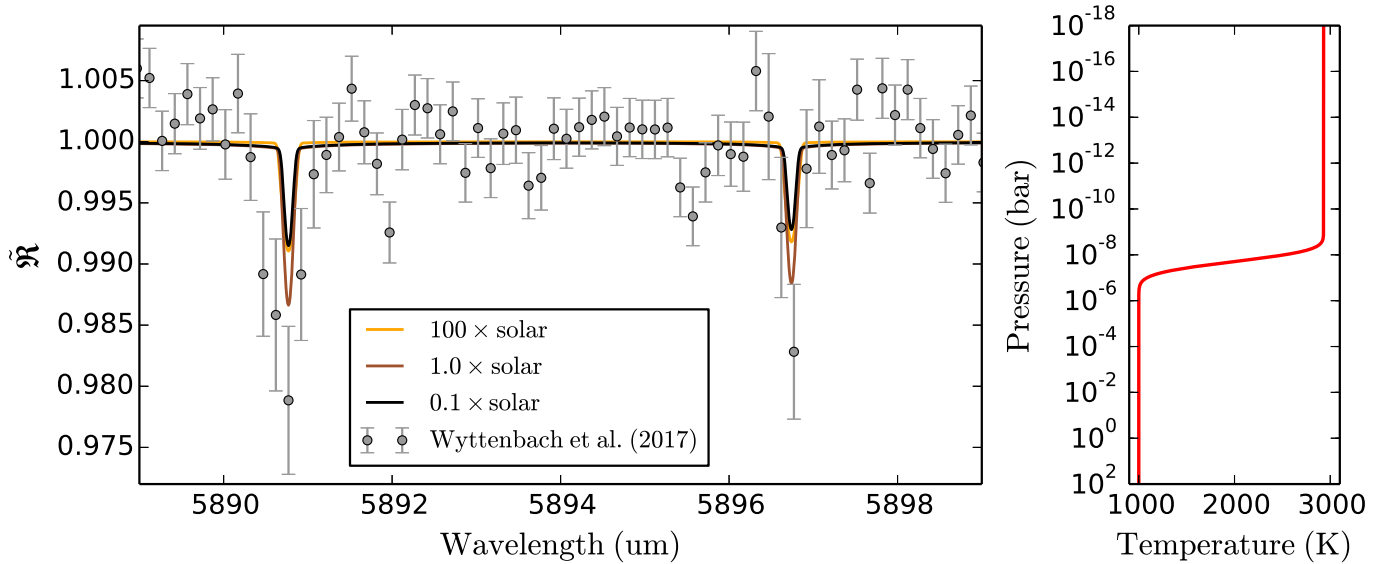}
\caption{{\bf Left:} WASP-49\,b high-resolution transmission spectra. The
  gray dots with $1\sigma$ error bar denote the reduced data
  from \citet{WyttenbachEtal2017aaWASP49bHARPS}, binned by 15 points.
  The solid curves denote our model spectra for each metallicity (see
  legend), adjusted to the reported spectral resolution of the data.
  {\bf Right:} atmospheric temperature profile adopted to compute the
  transmission models.}
\label{fig:hires}
\end{figure*}

\subsection{High-resolution Observations}
\label{sec:hires}

Recently, \citet{WyttenbachEtal2017aaWASP49bHARPS} published HARPS
high-resolution transmission observations of WASP-49\,b. They reported a
spectrally resolved detection of strong Na~{\small I} D lines, arising
from hot ($\sim$3000 K) high-altitude layers ($\sim$1.5 $R\sb{\rm p}$)
of the atmosphere, located above the layers studied in the previous
sections.  We decided not to include this dataset into our retrieval
analysis, because the high-resolution data requires a different data
reduction and modeling than that of the low-resolution data, beyond
the scope of this manuscript.

Essentially, ground-based high-resolution data analysis require a
double normalization approach where, in addition to the normalization
by the out-of-transit spectra, one needs to perform a spectral
normalization to remove (or reduce) time-correlated noise introduced
by telluric variations \citep[see section 4.1
of][]{WyttenbachEtal2017aaWASP49bHARPS}.  As a first look at this
problem, we present a forward-model approach for the simplified case
of isolated absorption lines arising above a wavelength-independent
cloud deck.
The Pyrat-Bay forward model returns the modulation spectrum:
\added{
\begin{equation}
M(\lambda) = \frac{f\sb{\rm out}(\lambda)-f\sb{\rm in}(\lambda)}
                  {f\sb{\rm out}(\lambda)},
\end{equation}
}
where $f\sb{\rm in}(\lambda)$ and $f\sb{\rm out}(\lambda)$ are the in-
and out-of-transit flux spectrum, respectively.  Then, $1-M$
corresponds to $f\sb{\rm in}(\lambda)/f\sb{\rm out}(\lambda)$.  If the
continuum is dominated by a gray absorber, we can choose any
wavelength $\lambda\sb{\rm ref}$ for the spectral normalization, as
long as the modulation spectrum at $\lambda\sb{\rm ref}$ is dominated
by the gray opacity.  Then we can compute,
\begin{equation}
\tilde{R} = 
   \frac{1-M(\lambda)}{1-M(\lambda\sb{\rm ref})} =
            \frac{f\sb{\rm in} (\lambda)/f\sb{\rm in} (\lambda\sb{\rm ref})}
                 {f\sb{\rm out}(\lambda)/f\sb{\rm out}(\lambda\sb{\rm ref})}.
\label{eq:rtilde}
\end{equation}

Equation (\ref{eq:rtilde}) effectively mimics the in- and
out-of-transit spectrum ratio in the planet rest frame,
$\tilde{\mathfrak{R}}$ \citep[equation 5
of][]{WyttenbachEtal2017aaWASP49bHARPS}.  Figure \ref{fig:hires} (left
panel) shows the reduced HARPS high-resolution data
from \citep{WyttenbachEtal2017aaWASP49bHARPS}, along with a selected
$\tilde{R}$ model for each of our sampled metallicities.
Like in the previous section, these models are in thermochemical and
hydrostatic equilibrium, but consider an extended hot upper atmosphere
at $\sim$3000~K (Fig. \ref{fig:hires}, right panel).

Our models agree with \citet{WyttenbachEtal2017aaWASP49bHARPS}, that
the Na signal is unexpectedly large.  Given the lower Na abundance of
the sub-solar case, and the smaller scale height of the super-solar
case, the solar-abundance case produces the largest Na signal, though
in each case our model underestimates the observed signal.  To match
the observed strength of the Na signal, we would need an enhanced Na
abundance, with respect to the other species.  Alternatively, at such
heights the atmosphere could not necessarily be in local thermodynamic
equilibrium.  Additionally, for any given metallicity, the Na signal
increases with a deeper aerosol layer, suggesting that the cloud deck
is located towards the lower end of the found posterior distribution.

With respect to the line width, even with an atmosphere at 3000~K, our
models underestimate the width of the Na signal (note that at these
altitudes, Doppler broadening dominates the line width).  This would
suggest that there are strong equatorial winds on the
planet \citep[e.g.,][]{LoudenWheatley2015apjHD189bWinds}.  In any
case, we regard these conclusions as tentative, since to obtain more
conclusive constraints from the high-resolution data require a more
detailed analysis, and possibly with a more elaborated Sodium
absorption profile model.

\section{CONCLUSIONS}
\label{sec:conclusions}

The presence of cloud condensates or photochemical hazes in the
atmosphere of exoplanets limits our ability to characterize these
atmospheres.  Aerosols not only hide molecular or atomic spectral
features from deeper layers in an atmosphere, but also pose a
challenge in distinguishing the compounds.
This study of WASP-49\,b shows that by combining a number of atmospheric
models, we can still constrain atmospheric properties of cloud-covered
planets, subject to the observational limitations of current
facilities.

Using hydrodynamic atmospheric models we estimate the penetration
depth of the stellar high-energy irradiation, located at
{\sim}$\ttt{-8}$ bar for WASP-49\,b (Fig.~\ref{fig:hydroruns}).  This
value determines the lowest possible pressure where aerosols could
exist, as \added{the high temperatures evaporate the aerosol
condensates and} high-energy stellar photons \added{can also
dissociate the} condensates.

Since we do not know the shape of the temperature profile for WASP-49\,b, we
estimate it from radiative-transfer forward-model runs in radiative
equilibrium.  From a series of clear and cloudy radiative-equilibrium
models, we found that this planet should have temperatures in the
800--1200~K range, above the 0.1~bar level (Fig.~\ref{fig:helios}).
Deeper layers are not relevant as they are not accessible by
transmission spectra, even in the most favorable case (low-metallicity
clear-atmosphere case, Fig.~\ref{fig:clear}).

Then, we retrieve the transmission photospheric pressure of the planet
$p\sb{\rm T}$ (that corresponding to the transit radius) with an MCMC
radiative-transfer run, constrained by the flat NIR transmission
spectrum of WASP-49\,b.  Modeling the aerosol layer as a gray,
constant-cross-section absorber we derive the marginal posterior
distribution of $p\sb{\rm T}$ for three metallicity scenarios, 100,
1.0, and 0.1 times solar elemental abundances, in thermochemical
equilibrium.  The 68\% HPD credible region of $p\sb{\rm T}$ is
restricted between {\sim}$\ttt{-3}$--$\ttt{-5}$~bar and the
{\sim}$\ttt{-7}$~bar levels (Fig.~\ref{fig:marginalZ}).  Increasing
the metallicity constrains $p\sb{\rm T}$ to the smaller region because
the aerosol layer needs to blanket the {\water} and K features, which
arise from higher layers in the atmosphere.  The upper boundary of
$p\sb{\rm T}$ at {\sim}$\ttt{-7}$~bar is ultimately determined by the
penetration depth of the high-energy stellar XUV flux.

Finally, considering that different aerosol compounds condense at
different pressures and temperatures, we investigate which are the
plausible condensates for each metallicity scenario.  Adopting a
temperature profile from the radiative-equilibrium models
({\sim}800--1000~K), we find a range candidates from sodium sulfide
(high metallicity, lower temperature), to alabandite or Cr (solar and
low metallicity), to Fe, forsterite, or enstatite (low metallicity).

There are a couple of considerations that could lead to stronger
constraints than those found here.  Planets with stronger stellar
high-energy fluxes should have deeper penetration depths, limiting the
upper aerosol boundary.  The vibrational mode of small sub-micron
sized condensates produce absorption features in the infrared, which
could help discern different cloud types, particles sizes, and
altitudes \citep{WakefordSing2015aaHotJupiterClouds}.
Clearly, constraining cloudy-atmospheres' properties is heavily
limited by both the current data quality and the physical properties
of condensates, this exercise is no exception.  However, by adopting a
number of reasonable assumptions and combining multiple atmospheric
models, \added{we showed that one can constrain the location of an aerosol
layer on cloudy exoplanets with current data.}
As more and better-quality data becomes available, these and
other theoretical studies will help us to better characterize
exoplanet atmospheres, even in the case of cloudy skies.

\acknowledgments

We thank Dr. Kevin Heng for constructive discussions.  We thank
contributors to Numpy \citep{vanderWaltEtal2011numpy}, SciPy
\citep{JonesEtal2001scipy}, Matplotlib
\citep{Hunter2007ieeeMatplotlib}, the Python Programming Language, and
the free and open-source community; the
developers of the aastex latex template \citep{AASteam2016aastex61};
the NASA Astrophysics Data System;
and the JPL Solar System Dynamics group for software and services.  We
thank the anonymous referee for comments that improved the quality the
paper.  We acknowledge the Austrian
Forschungsf{\"o}rderungsgesellschaft FFG projects ``RASEN'' P847963
and ``TAPAS4CHEOPS'' P853993, the Austrian Science Fund (FWF) NFN
projects S11607-N16 and S11604-N16, and the FWF project P27256-N27.
Part of this work has been carried out within the frame of the
National Centre for Competence in Research ``PlanetS'' supported by
the Swiss National Science Foundation (SNSF). A.W. acknowledge
financial support of the SNSF, grants 200020\_152721 and
200020\_166227.  This article is based on photometric observations
made with FORS2 on the ESO VLT/UT1 (Prog. ID 090.C-0758), and the ESO
3.6 m telescope at the La Silla Observatory (ESO Prog. 096.C-0331).
The Pyrat-Bay Reproducible
Research Compendium (RRC) of this article will be available at
\href{https://github.com/pcubillos/CubillosEtal2017\_WASP49b}
{https://github.com/pcubillos/CubillosEtal2017\_WASP49b}
once the Pyrat-Bay code gets released for public use.

\software{Pyrat Bay: Python Radiative Transfer in a Bayesian framework
 (\href{http://pcubillos.github.io/pyratbay}
       {http://pcubillos.github.io/pyratbay}), HELIOS
 (\href{https://github.com/exoclime/HELIOS}
       {https://github.com/exoclime/HELIOS}) and latex template
 (\href{https://github.com/pcubillos/ApJtemplate}
       {https://github.com/pcubillos/ApJtemplate}).  }

%\bibliography{w49b_cloudy}

\end{document}